# Ultrafast Manipulation of a Double Quantum Dot via Lyapunov Control Method

Shuang Cong, *Senior Member, IEEE*, Ming-yong Gao, Long-zhen Hu，Guo-ping Guo, Gang Cao, Guang-can Guo

*Abstract*—For a double quantum dot (DQD) system, here we propose alternative ultrafast manipulate approach: Lyapunov control method, to transfer the state from $|R\rangle$ to $|L\rangle$ on the picosecond scale, orders of magnitude faster and transfer probability higher than the previously measured electrically controlled charge- or spin-based quits. The control laws are composed of two-direction components, one is used to eliminate the dissipation in the system, another is used to transfer the state. The control theory's stability ensures the system can be transferred to the target state in high probability, and the coefficients in control laws leads very fast convergence. The role of eliminating the dissipation plays the suppression of decoherence effect. Numerical simulation results show that under the realistic implementation conditions, the transfer probability and fidelity can be increased up to 98.79% and 98.97%, respectively. This is the first result directly applicable to a DQD system's state transferring using the Lyapunov control method. We also give specific experimental realization scheme.

## I. Introduction

SEMICONDUCTOR quantum dot is an artificial solid-state quantum system, its shape and size are manageable, easy to manipulate and measure, moreover, it can make use of the mature semiconductor integrated circuit technology in classical computer, all of these advantages make the quantum dots be highly scalable and become one of the powerful candidates for quantum compute. The decoherence time of the free electrons in semiconductor quantum dots is usually within a few nanoseconds ($10^{-9}s$) [1-3], the control operation on the picoseconds ($10^{-12}s$) timescale may be necessary, namely the key point of the qubit manipulation need to accomplish the manipulation process of the quantum system before the decoherence. In recent years, people have worked intensively on the experimental apparatus for realizing this goal. LZS interference was first proposed by Landau, Zener and Stückelberg which occurs when the control field sweeps through the anti-crossing of a two-level system. There will generate a significant tunneling from the ground state to the excited state, in this way, the interference caused by the evolutionary trajectory of ground state and exited state will lead to the LZS interference. As the target state is defined by the constructive interference of LZS interference, and research results show that the LZS interference method is robust to certain types of noise and might enable the implementation of manipulating qubits with high fidelity [4-7], therefore the coherent dynamics of LZS interference process aroused a great deal of interest for quantum control [8]. Cao *et. al.* carried out the qubits state transfer in a double quantum dot (DQD) system by utilizing the LZS interference [9]. In their experiments, the system could transfer from the initial state to target state with a probability of ~ 68%, and the fidelity of the system could reach ~ 80%.

Can other control methods be used for further improving the state transfer performances of the DQD system? Our answer is yes.

The manipulation of the dynamics characteristic of the quantum system is intended to design a suitable control strategy on one or more performance indexes, which can lead the system evolving to the desired target state with high probability [10]. Quantum Lyapunov control method has been studied for ten years, and obtained series of research achievements [11, 12]. We expect to use quantum Lyapunov control method [13] to design a Lyapunov control field specially for the state transfer of two-level DQD system in order to obtain higher state transfer performances. This paper will theoretically provide a more ultrafast control method for two-level DQD system. We verify the superiorities of the designed Lyapunov control field in state transfer probability by numerical simulations.

## II. Establishment of two-level DQD system Model based on the LZS interference

A Two-level DQD system is prepared within a GaAs/AlGaAs hetero structure, Fig. 1 illustrates the schematic diagram of the two-level DQD system, in which

Manuscript received Jan. 12, 2014. This work was supported in part by the National Key Basic Research Program under Grant No. 2011CBA00200.

S. Cong is with the Department of Automation, University of Science and Technology of China, Hefei 230027, China, phone: 86-551-63602224; fax: 86-551-63603244; e-mail: scong@ustc.edu.cn.

M. Y. Gao, and Long-zhen Hu are with the Department of Automation, University of Science and Technology of China, Hefei 230027, China

G.P Guo, G. Cao, L. Wang, G.C. Guo are Key Laboratory of Quantum Information, University of Science and Technology of China, Chinese Academy of Sciences, Hefei 230026, China, phone: 86-551-63606043; fax: 86-551-63606828; e-mail: gpguo@ ustc.edu.cn).

$A_1$ to $A_4$ and $B_1$ to $B_4$ are 8 metal gate patterns; $C_1$ and $C_2$ are two ohmic contacts; $M$ is an external conductivity detector, $QPC$ is quantum point contact; two quantum dots are formed substantially at position presented by $L$ and $R$. The gate patters of $A_1$ and $B_1$ control the coupling strength between quantum dots. $A_2$ and $B_2$ separate the quantum dots from the $QPC$. $A_3$ and $B_3$ are the plunger gate of two quantum dots, on which the control fields impose to manipulate the dynamic characteristics of quantum dots. $A_4$ and $B_4$ control the channel current of the $QPC$. The opening upon $L$ and $R$ is used to enhance the coupling between the two dots and increase the sensitivity of the $QPC$. Every change of the number of electrons in quantum dots will lead to the change of conductance in the $QPC$, denoted as $G_{QPC}$. The $G_{QPC}$ represents the state change of the dots and can be measured by $M$, in such a way, there is no need to measure the change of the current in dots and the difficulty for measuring is avoided.

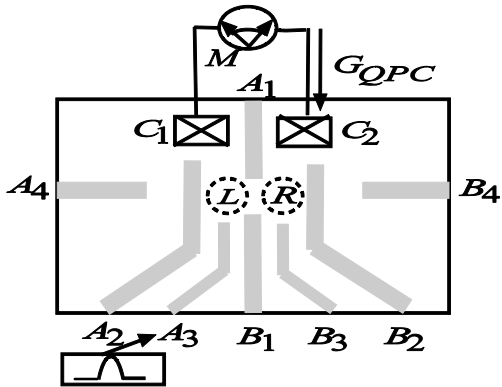

Fig.1 Schematic diagram of the of two-level DQD system

Fig, 2 is two-level DQD system state transfer diagram which $E_L$ and $E_R$ are the energy levels for an electron in left and right quantum dots $L$ and $R$, respectively. The size of $E_L$ can be adjusted by applying external control field at the gate pattern $A_3$, similarly $E_R$ can be adjusted by applying external control field at the gate pattern $B_3$. The system detuning $\varepsilon$ is denoted by the difference of $E_R$ and $E_L$:

$$\varepsilon = E_R - E_L \quad (1)$$

According to the LZS interference, the key to the state transfer of two-level DQD system is adjusting the system detuning $\varepsilon$. One can see from Eq. (1) that the energy levels $E_R$ and $E_L$ can be adjusted by applying external control field, thus the level detuning $\varepsilon$ is adjustable by external control field. Next, we'll focus on deriving the relation between the state transfer probability and the system detuning $\varepsilon$. When the external control field takes the system to the anti-crossing ($\varepsilon = 0$), the Landau-Zener tunneling will take place in the two-level DQD system, the ground state will jump into the excited state with the probability of $P_{LZ}$, as the external control field takes the system far away from the anti-crossing, the two energy level of ground and excited states will come up Stücklberg interference. When the control field takes the system to the anti-crossing the second time, the accumulated phase caused by the interference of two energylevels is $\phi_i$. The constructive interference occurs when the accumulated phase $\phi_i = 2N\pi$, $N = 0,1,2...$, the system will transfer from initial state to desired state. The destructive interference occurs when the accumulated phase $\phi_i = 2N\pi + \pi/2$, $N = 0,1,2...$ the system will remain at the initial state.

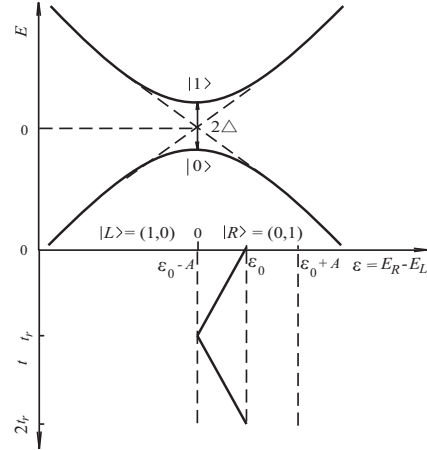

Fig, 2 Two-level DQD system state transfer diagram

The state transfer probability from initial state to desired state throughout the process is [7]:

$$P_{|L\rangle} = 2P_{LZ}(1 - P_{LZ})\cos(2\phi_{LZ} - \phi_i) \quad (2)$$

where, $\phi_{LZ}$ is a phase related to Stokes phenomenon which is small enough to be ignored [9].

The formulas of $P_{LZ}$ and $\phi_i$ are as follows:

$$P_{LZ} = \exp(-2\operatorname{Im}\int_0^{t_0}[E_+(t') - E_-(t')]dt) \quad (3)$$

$$\phi_i = \int_0^{t_r}[E_+(t) - E_-(t)]dt + \int_{t_r}^{2t_r}[E_+(t) - E_-(t)]dt \quad (4)$$

where, $E_+$ and $E_-$ represent the energy levels of the bonding and anti-bonding states, respectively. The relation between the detuning $\varepsilon$ and $E_\pm$ is: $E_\pm = \pm\sqrt{\varepsilon^2 + \Delta^2}$, where $\Delta$ is the energy level difference of the bonding and anti-bonding states at the anti-crossing, denoted as anti-crossing gap.

The control field to realize the state transfer of quantum dots based on LZS interference is a periodic function with a amplitude of $A$. Cao $et.\ al.$ adopted a triangular periodic function with the amplitude of $A$ in their experiment [9], in which the rising time is denoted as $t_r$, the cycle of the control field is $2t_r$ as shown in Fig. 2. Under their control field designed, the relation between the detuning $\varepsilon$ and external control field is [9]:

$$\varepsilon(t) = \begin{cases} \varepsilon_0 - vt & 0 < t \le t_r \\ \varepsilon_0 - A + vt & t_r < t \le 2t_r \end{cases} \quad (5)$$

where, $v = A/t_r$ represents the rising velocity of the control field, $\varepsilon_0$ is initial value of the detuning generated in the preparation process of two-level DQD system.

Take $E_\pm = \pm\sqrt{\varepsilon^2 + \Delta^2}$, and Eq. (5) into Eq. (3) and Eq. (4), respectively, and compute the integration, one can obtain:

$$P_{LZ} = \exp(-\frac{2\pi\Delta^2}{v\hbar}) \quad (6)$$

$$\phi_i = \frac{2(A-\varepsilon_0)^2}{v\hbar} \quad (7)$$

where, $\hbar = h/2\pi$.

Take the Eq. (6) Eq. (7) and $v = A/t_r$ into the transfer probability Eq. (2), one can obtain:

$$P_{|L\rangle} = 2\exp(-\frac{2\pi\Delta^2 t_r}{A\hbar})[1-\exp(-\frac{2\pi\Delta^2 t_r}{A\hbar})]\cos[2\frac{(A-\varepsilon_0)^2 t_r}{A\hbar}] \quad (8)$$

The action time of control field is taken one cycle in the experiment, namely: $t_p = 2t_r$, taking it into the Eq. (8) one can obtain the relation of the transfer probability $P_{|L\rangle}$ with the control field amplitude $A$ and the action time $t_p$ as:

$$P_{|L\rangle} = 2\exp(-\frac{\pi\Delta^2 t_p}{A\hbar})[1-\exp(-\frac{\pi\Delta^2 t_p}{A\hbar})]\cos[\frac{(A-\varepsilon_0)^2 t_p}{A\hbar}] \quad (9)$$

which is an indirect relation between the detuning $\varepsilon$ and the transfer probability $P_{|L\rangle}$ and the relationship of $v = A/t_r$, $t_p = 2t_r$. Eq. (9) is also the relationship used in the experiment for adjusting the system state transfer directly.

According to the relationship of control field and transfer probability in Eq. (9), and the relation between the detuning $\varepsilon$ and external control field in Eq. (5), one can obtain the Hamiltonian of two-level DQD system as:

$$\begin{aligned} H &= \frac{1}{2}\varepsilon(t)\sigma_z + \Delta\sigma_x = \frac{1}{2}\varepsilon_0\sigma_z + \frac{1}{2}f(t)\sigma_z + \Delta\sigma_x \\ &= H_0 + H_c \end{aligned} \quad (10)$$

where, $f(t) = \begin{cases} -vt & 0 < t \le t_r \\ -A+vt & t_r < t \le 2t_r \end{cases}$, $\sigma_x$ and $\sigma_z$

represent the Pauli matrices: $\sigma_x = \begin{pmatrix} 0 & 1 \\ 1 & 0 \end{pmatrix}$, $\sigma_z = \begin{pmatrix} 1 & 0 \\ 0 & -1 \end{pmatrix}$,

$H_0 = \frac{1}{2}\varepsilon_0\sigma_z$, $H_c = \frac{1}{2}f(t)\sigma_z + \Delta\sigma_x$.

In next section, we will make use of the Markov master equation, set out in terms of the system control theory, and study the control field of state transfer probability.

## III. DESIGN OF CONTROL FIELD FOR THE TWO-LEVEL DQD SYSTEM BASED ON LYAPUNOV CONTROL METHOD

In this section, we design a control field for two-level DQD system based on Lyapunov control method in order to obtain higher transfer probability and shorter time from the initial state to the desired state within decoherence time.

The master equation of the controlled two-level DQD system is [9]:

$$\frac{d\rho}{dt} = -\frac{i}{\hbar}[H,\rho] + L \quad (11)$$

where, $H = H_0 + H_c$, $H_0 = (1/2)\varepsilon_0\sigma_z$, $H_c = \sum_{m=1}^{2} f_m H_m$, $L$ is the dissipation described by standard Lindblad form:

$$L = L_1\rho L_1^\dagger + L_2\rho L_2^\dagger - \frac{1}{2}(\rho, L_1^\dagger L_1) - \frac{1}{2}(\rho, L_2^\dagger L_2) \quad (12)$$

where, $L_1 = \sqrt{\Gamma_1}\sigma_-$, $L_2 = \sqrt{\Gamma_2}\sigma_z$, $\Gamma$ represents the decoherence rate, and $\Gamma_1 = 1/T_1$, $\Gamma_2 = 1/T_2$, $T_1$ and $T_2$ represent the decoherence time of two quantum dots, respectively.

The Lyapunov control method is a designing control law method by means of the Lyapunov indirect stability theorem. This theorem was originally used to judge if a system was stable, and later it was widely used to design at least a stable control system. Now the Lyapunov control method becomes a very popular control law design method in systems control community like the optimal control method, and it has the advantages of easy to design, and the control law has analytical form [11]. For the design of the control law based on Lyapunov control method, one should select a Lyapunov function $V(x)$ which is semi-definite positive and differentiable in the phase space $\Omega = (x)$. The control law of the system is acquired by ensuring the system stable condition: $\dot{V}(x) \le 0$. Therefore, the key point to design the control law is to find a suitable Lyapunov function $V(x)$. There are many candidate Lyapunov functions. Here we chose the Lyapunov function based on state distance as [13]:

$$V = \frac{1}{2}tr\left((\rho - \rho_f)^2\right) \quad (13)$$

where, $\rho$ represents system state, $\rho_f$ represents final state.

In order to obtain the control law that ensures the stability of the system, one can solve the first time derivative of Eq. (13) and obtain:

$$\begin{aligned}
\dot{V} &= tr\left(\dot{\rho}(\rho-\rho_f)\right) = tr\left(\left(-\frac{i}{\hbar}[H,\rho]+L\right)(\rho-\rho_f)\right) \\
&= tr\left(\left(-\frac{i}{\hbar}\left[H_0+\sum_{m=1}^{2}f_m(t)H_m,\rho\right]+L\right)(\rho-\rho_f)\right) \\
&= tr\left(\left(-\frac{i}{\hbar}[H_0,\rho]+L\right)(\rho-\rho_f)\right) + tr\left(-\frac{i}{\hbar}\sum_{m=1}^{2}f_m(t)[H_m,\rho](\rho-\rho_f)\right) \\
&= f_1(t)\cdot D_1 + f_2\cdot D_2 + C
\end{aligned}$$

(14)

where, $D_m = tr\left(-\frac{i}{\hbar}[H_m,\rho](\rho-\rho_f)\right)$ ($m = 1, 2$) is a real function of $\rho$, $f_m(t)$ ($m = 1, 2$) is the control law to be (14) solved, $C = tr\left(\left(-\frac{i}{\hbar}[H_0,\rho]+L\right)(\rho-\rho_f)\right)$ is the dissipation term of the system.

Because the symbol of the dissipation term $C$ in Eq. (14) may be positive and negative, it could lead the symbol of $\dot{V}(x)$ to be uncertain. In order to obtain the control laws that ensure $\dot{V}(x)\leq 0$, the main idea of designing the control laws is that we adopt two control fields, one control is used to offset the influence of the dissipation term $C$, another is used to manipulate the state transfer. With the coordination of the two control fields, the $\dot{V}(x)\leq 0$ can always be hold during the whole control procedure. On the other hand, an adjustable threshold variable $\theta$ is introduced to limit the value of $D_m$, and used to determine which control of two to counteract the dissipation term. The specific design process is as follows:

1) In Eq. (4), if $|D_1|>\theta$ holds, then we design the control laws as: $f_1 = -C/D_1$, which is to offset $C$, and choose $f_2 = -g_2\cdot D_2$. In this way, Eq. (14) becomes $\dot{V} = -g_2\cdot D_2^2 \leq 0$. Then the control laws can be written as: $f_1 = -C/D_1$, $f_2 = -g_2\cdot D_2$, and $g_2 > 0$.

2) In Eq. (14), if $|D_1|<\theta$ and $|D_2|>\theta$ hold, then we design $f_2$ to counteract the dissipation term $C$. Similar to 1), the designed control laws can be written as: $f_1 = -g_1\cdot D_1$, $g_1 > 0$, and $f_2 = -C/D_2$ is used to adjust the control amplitude and ensure $\dot{V} = -g_1\cdot D_1^2 \leq 0$.

3) In Eq. (14), if $|D_1|<\theta$ and $|D_2|<\theta$ hold, then we calculate the value of the Lyapunov function $V$ to estimate the distance between the controlled state and the target state. The control object is deemed to be achieved if the transfer error has reached a pre-given value $\varepsilon$, otherwise one needs to reselect the control parameters $g_1$ and $g_2$.

According to the analysis mentioned above, the control fields $f$ are designed as:

$$f = \begin{cases} f_1 = -C/D_1 \\ f_2 = -g_2\cdot D_2 \end{cases} \quad |D_1|>\theta$$

$$f = \begin{cases} f_1 = -g_1\cdot D_1 \\ f_2 = -C/D_2 \end{cases} \quad |D_1|<\theta, |D_2|>\theta$$

(15)

where, $\theta$ is the threshold introduced to avoid the mathematical singularity in the fractional expression of the control laws. We set $\theta = 5\times 10^{-6}$. $g_1$ and $g_2$ are the adjustable parameters of the control fields.

IV. NUMERICAL SIMULATION AND RESULT ANALYSIS

To investigate the performances of the control methods proposed in this paper, we have done two things as follows: 1) Numerical simulation of the two-level DQD system state transfer based on the LZS interference method, and result analysis. 2) Numerical simulation of state transfer based on quantum Lyapunov control method proposed, and the results comparison with the LZS interference method.

In the control field (15), the most important thing is to determine the two adjustable control parameters $g_1$ and $g_2$, which determine the amplitude of the control law (15) at each sample time and also decide the performance of the control system. In our simulation the density matrix in the control is obtained by numerically solving the master Eq. (11). One of the numerical simulations' purposes is to acquire the optimal parameters of $g_1$ and $g_2$ in the different system's parameters.

The qubit transfer probability and ultrafast control performance of the Lyapunov control are studied in different control parameters. At the detuning energy of $\varepsilon_0 = 400$ $\mu$ev, the control amplitudes are limited in the 800 $\mu$ev. The decoherence time $T_2 = 5$ ns, the spin relaxation time $T_1 = 5$ ns, the time-resolution is 0.1 ps, the simulations of three groups control parameters of $g_1$ and $g_2$, their transfer probabilities, and the control laws as functions with time duration $t_p$ are shown in Fig. 1, from which one can see that the qubit transfer probability can achieve 99.95% at $t_p = 50$ ps with $g_1 = g_2 = 1$. This simulation results indicate that the Lyapunov control is an ultrafast control, and at the same time it has the very high state transfer probability which is guaranteed by the Lyapunov stability theorem.

Considering the external control field will be generated in the Lab. by the Agilent 81134A pulse generator with the time-resolution of 1 ps [9], the realistic parameters of the Lyapunov control are further studied. To obtain the parameters with good performance, the transfer probability $P_{|L\rangle}$ as a function of adjustable parameter $G$ ($g_1 = g_2$) and the initial detuning $\varepsilon_0$ is firstly studied with the range of $G$ in $[0.1, 2]$ and the energy position $\varepsilon_0$ in $[0, 400]$ $\mu$ev, and then we refine the ranges of parameters and the simulation results of the charge qubit dynamics using the realistic control fields are shown as in Fig. 2a, where the range of the adjustable parameter $G$ is $[0.12, 0.24]$ and the energy position $\varepsilon_0$ is $[0, 100]$ $\mu$ev. In this way we find a better adjustable control parameters $g_1 = g_2 = 0.22$. Fixed these control parameters, the transfer probability $P_{|L\rangle}$ as a function of the energy position $\varepsilon_0$ and the control duration time $t_p$ is also studied, which is shown as in Fig. 2b. The better energy position $\varepsilon_0 = 90$ $\mu$ev is obtained according to the simulations, which can result in a higher transfer probability of the qubit and a fast rising time of the coherent oscillation. The control fields in the DQD system under the above realistic parameters obtained are shown as in Fig. 2c. In fact the control function used to against the dissipation effect plays the rule of extending the decoherence time $T_2$ according to the idea introduced in the spin-echo method [14]. Compared with the Gaussian-shaped short pulse used as the LZS interference [9], the shapes of our Lyapunov control fields are the optimal-shaped function pulses.

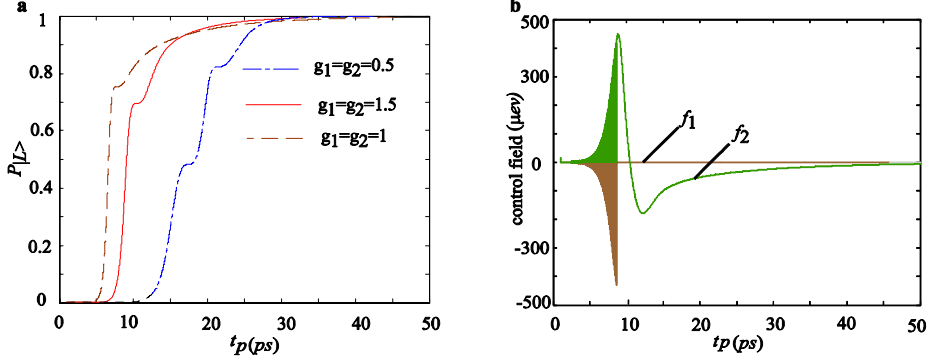

Fig. 1 Transfer probabilities and control laws, which displays a) transfer probabilities with $g_1 = g_2 = 0.5$, $g_1 = g_2 = 1$, and $g_1 = g_2 = 1.5$, and b) Control laws as functions with the pulse duration time $t_p$ with the adjustable parameters $g_1 = g_2 = 1$.

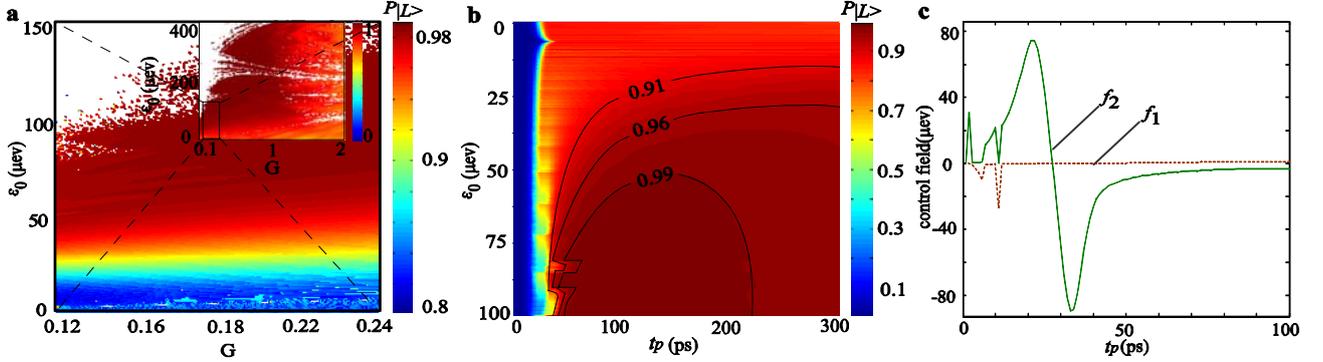

Fig. 2 Simulation results of the charge qubit dynamics using the realistic control fields, which displays a) the probability of the qubit in state $|L\rangle$ as a function of energy position $\varepsilon_0$ and the control parameter G ($g_1 = g_2$), in which the range of energy position $\varepsilon_0$ is in [0, 150] $\mu$ev; the inset shows the range of $\varepsilon_0$ in [0, 400] $\mu$ev; the white area represents the control method can not effectively drive the qubit transfer under the specific parameters of $\varepsilon_0$ with $g_1 = g_2 = 0.22$ and $G$ with the time-resolution of 1ps. b) the transfer probability $P_{|L\rangle}$ as a function of the energy position $\varepsilon_0$ and the pulse duration time $t_p$, one can see that at the energy position of $\varepsilon_0 = 90$ $\mu$ev, the state of $|L\rangle$ has a fast rising time and high probability. c) Lyapunov control fields with $g_1 = g_2 = 0.22$, in which control function $f_2$ is used to drive the state transfer, control field $f_1$ is used to eliminate the dissipation existed in the system.

The density matrices of the qubit state transfer for the DQD system from the initial state $|R\rangle$ to the final state $|L\rangle$ are shown in Fig. 3, in which $\rho_{11}$ is probability of the qubit in state $|L\rangle$. The maximum transfer probability ($\rho_{11}$ in Fig. 3a) under Lyapunov control is 98.79%, which appears at $t_p = 71$ ps. We also carried out the simulation performance of the density matrix under the LZS interference introduced in [9], and the result shows that the transfer probability ($\rho_{11}$ in Fig. 3b) is 67.68% at $t_p = 116$ ps, which fits the results of the experiments in [9] perfectly.

V. PERFORMANCE ANALYSIS

The Bures fidelity is used in the numerical simulations, which is defined as [15][16]:

$$F = Tr\left[\sqrt{\rho_f}\rho_s\sqrt{\rho_f}\right] \quad (16)$$

where $\rho_s$ is the density matrix of the system, $\rho_f$ is the final density matrix, and

$$\rho_f = |L\rangle\langle L| = (0\ 1)^T(0\ 1) = \begin{pmatrix} 0 & 0 \\ 0 & 1 \end{pmatrix}.$$

The simulation results of operation fidelity defined in (16) as a function of the pulse duration time $t_p$ and decoherence time $T_2$ are shown in Fig. 4a. These results reveal that given an appropriate regime of charge-state decoherence, at the control time of ~ 100 ps, the value of the fidelity higher than 98.97% ($T_2 = 5$ns) can be achieved under the Lyapunov control. The performance of the system fidelity has great advantage over that based on the LZS interference not only in the robustness against the dissipation, but also in the fidelity.

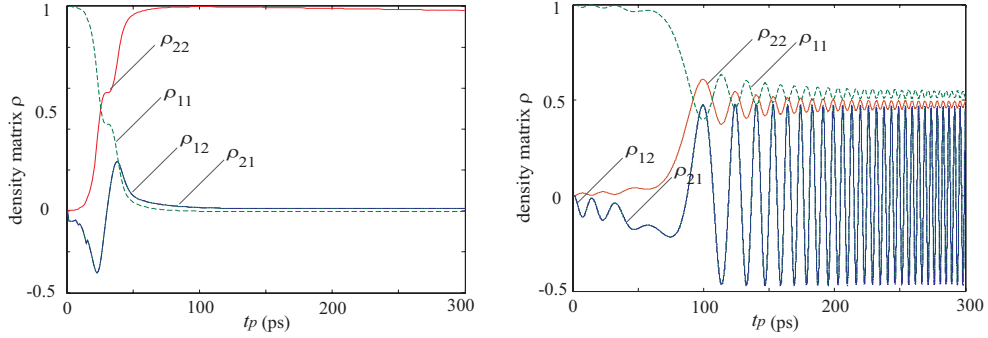

Fig. 3 Simulation results of the state transfer for the DQD system. (a) State transfer probability $P_{|L\rangle} = \rho_{11}$ under the Lyapunov control. (b) State transfer probability $P_{|L\rangle} = \rho_{11}$ under the LZS interference pattern.

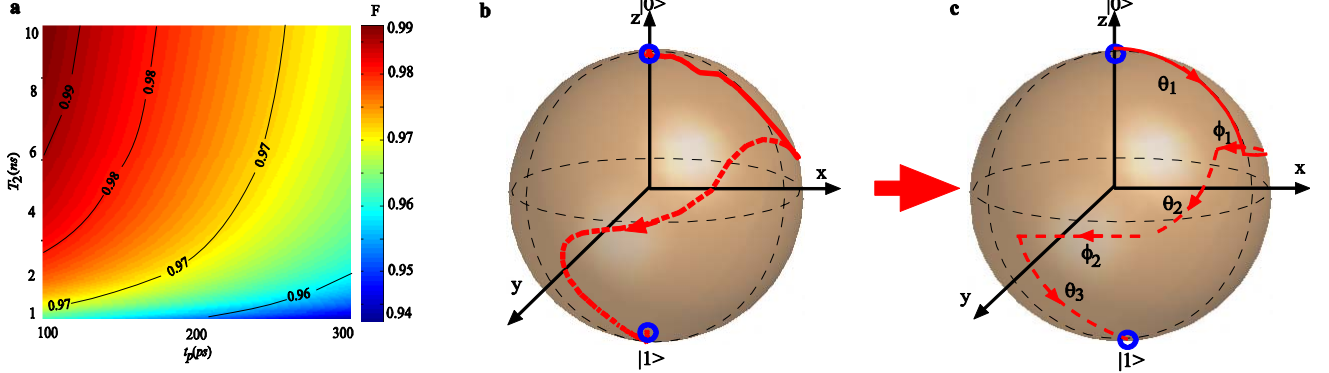

Fig.4 Simulation results, which displays a) the fidelity as a function of the pulse duration time $t_p$ and decoherence time $T_2$ under the Lyapunov control method in $\varepsilon_0 = 90\mu ev$, $g_1 = g_2 = 0.22$, $T_1 = 5$ ns; b) the qubit transfer trajectory on the Bloch sphere; c) the schematic diagram to disintegrate the trajectory into unitary operation matrices on the Bloch sphere.

The Hamiltonian of a two-level double quantum dot under the LZS interference control is [9]:

$$H = \frac{1}{2}\varepsilon(t)\sigma_z + \Delta\sigma_x \quad (17)$$

where $\varepsilon(t) = \begin{cases} \varepsilon_0 - vt, & 0 < t < t_r \\ \varepsilon_0 - A + v(t-t_r), & t_r < t < 2t_r \end{cases}$.

$$= \frac{1}{2}\varepsilon_0\sigma_z + \Delta\sigma_x + \frac{1}{2}f(t)\sigma_z$$

We rewrite this Hamiltonian of the system as:

$$H = \frac{1}{2}\varepsilon_0\sigma_z + \left[\Delta\sigma_x + \frac{1}{2}f(t)\sigma_z\right]$$
$$= \frac{1}{2}\varepsilon_0\sigma_z + \left[f_1^{LZS}(t)\sigma_x + \frac{1}{2}f_2^{LZS}(t)\sigma_z\right],$$
$$= H_0^{LZS} + H_c^{LZS}$$

where, $H_0^{LZS} = \frac{1}{2}\varepsilon_0\sigma_z$,

$H_c^{LZS} = f_1^{LZS}\sigma_x + \frac{1}{2}f_2^{LZS}\sigma_z$,

$f_1^{LZS}(t) = \Delta$,

$f_2^{LZS}(t) = \begin{cases} -vt, & 0 < t < t_r \\ -A + v(t-t_r), & t_r < t < 2t_r \end{cases}$.

In the same way, Hamiltonian of the system under the Lyapunov control fields can be written as:

$$H = H_0^{LY} + H_c^{LY} = \frac{1}{2}\varepsilon_0\sigma_z + f_1^{LY}\sigma_x + f_2^{LY}\sigma_y$$

where, $H_0^{LY} = \frac{1}{2}\varepsilon_0 \sigma_z$, $H_c^{LY} = f_1^{LY}\sigma_x + f_2^{LY}\sigma_y$;

when $|D_1| > \theta$, one has

$$f_1^{LY}(t) = -\frac{tr\left(\left(-\frac{i}{\hbar}[H_0,\rho(t)]+L\right)(\rho(t)-\rho_f)\right)}{tr\left(-\frac{i}{\hbar}[H_1,\rho(t)](\rho(t)-\rho_f)\right)}, \text{ and}$$

$$f_2^{LY}(t) = -g_2 \cdot tr\left(-\frac{i}{\hbar}[H_2,\rho(t)](\rho(t)-\rho_f)\right);$$

when $|D_1| < \theta$ and $|D_1| > \theta$, one has

$$f_1^{LY}(t) = -g_1 \cdot tr\left(-\frac{i}{\hbar}[H_1,\rho(t)](\rho(t)-\rho_f)\right), \text{ and}$$

$$f_2^{LY}(t) = -\frac{tr\left(\left(-\frac{i}{\hbar}[H_0,\rho(t)]+L\right)(\rho(t)-\rho_f)\right)}{tr\left(-\frac{i}{\hbar}[H_2,\rho(t)](\rho(t)-\rho_f)\right)}.$$

So in the Lyapunov control pattern, the directions of the control field $f_1^{LY}$ in $H_c^{LY}$ and $f_1^{LZS}$ in $H_c^{LZS}$ in the LZS interference pattern are both applied in the orientation of $\sigma_x$. The difference between the control field $f_2^{LY}$ in $H_c^{LY}$ and $f_2^{LZS}$ in $H_c^{LZS}$ is: $f_2^{LY}$ is applied in the $\sigma_y$, while $f_2^{LZS}$ is in the $\sigma_z$. The Bloch sphere model provides a convenient picture to understand the Lyapunov control of the charge qubit, in which the ground state and the exited state $|R\rangle$ and $|L\rangle$ are represented by $|0\rangle$ and $|1\rangle$, respectively. The dynamics of the qubit can be represented by applying the appropriate sequence of unitary operation matrices to the initial state. The matrices

$$R_y(\theta) = \exp(-i\theta\sigma_y/2),$$
$$R_z(\phi) = \exp(-i\phi\sigma_z/2),$$

give the rise to a rotation on the Bloch sphere around the $y$ axis by angle $\theta$ and around $z$ axis by an angle $\phi$.

The qubit transfer trajectory on the Bloch sphere under Lyapunov control shown in Fig. 4b can be explained by the schematic diagram to disintegrate the trajectory into unitary operation matrices on the Bloch sphere shown in Fig. 4c, in which the charge qubit is initiated at state $|0\rangle$, and firstly the control field $f_2^{LY}$ rotates the state around $y$ axis by angle $\theta_1$, then the energy detuning rotates the state around $z$ axis by angle $\phi_1$, then after the control field $f_2^{LY}$ again rotates the state around $y$ axis by angle $\theta_2$, afterwards the detuning rotates the state around $z$ axis by angle $\phi_2$, in the end the control field $f_2^{LY}$ rotates the state around $y$ axis by angle $\theta_3$. In the whole process, the control field $f_1^{LY}$ contributes little rotation on the Bloch sphere because it is designed to eliminate the dissipation, in this way the qubit state can be preserved on the surface of the Bloch sphere. With these means, the qubit can be effectively driven from the state $|0\rangle$ to the state $|1\rangle$.

## VI. CONCLUSION

This paper studied the control field based on LZS interference firstly, designed the control field based on quantum Lyapunov control method, and carried out the numerical simulations. The performances of the state transfer probability of the two-level DQD system are analyzed and compared. The work of this paper is the first time to successfully manipulate the state transfer of two-level DQD system via Lyapunov control method, which can be realized in the actual two-level DQD system experiments with high possibility.

## BIOGRAPHIES

**Shuang Cong** received a B.S. degree in Automatic Control from Beijing University of Aeronautics and Astronautics in 1982, and a Ph.D. degree in System Engineering from the University of Rome "La Sapienza" Rome, Italy, in 1995. She is currently a professor in the Department of Automation at the University of Science and Technology of China. Her research interests include advanced control strategies for motion control, fuzzy logic control, neural networks design and applications, robotic coordination control, and quantum system control.

## REFERENCES

[1] Petersson, K. D., Petta, J. R., Lu, H. & Gossard, A. C. Quantum coherence in a one-electron semiconductor charge qubits. Phys. Rev. Lett. 105, 2010, 246804.
[2] Nowack, K. C., Koppens, F. H. L., Nazarov, Y. V. & Vandersypen, L. M. K.Coherent control of a single electron spin with electric fields. Science 318, 2007, pp. 1430–1433.
[3] Petta, J. R., Lu, H. & Gossard, A. C. A coherent beam splitter for electronic spin states. Science 327, 2010, pp. 669–672.
[4] Berezovsky, J., Mikkelsen, M. H., Stoltz, N. G., Coldren, L. A. & Awschalom, D. D.Picosecond coherent optical manipulation of a single electron spin in a quantum dot. Science 320, 2008, pp. 349–352.
[5] Press, D., Ladd, T. D., Zhang, B. & Yamamoto, Y. Complete quantum control of a single quantum dot spin using ultrafast optical pulses. Nature 456, 2008, pp. 218–221.
[6] De Greve, K. et al. Ultrafast coherent control and suppressed nuclear feedbackof a single quantum dot hole qubits. Nat. Phys. 7, 2011, pp. 872–878.
[7] Shevchenko, S. N., Ashhab, S. & Nori, F. Landau-Zener-Stückerlberg interferometry. Phys. Rep. 492, 1–30 (2010).
[8] Mooij H Mach-Zehnder Interferometry in a Strongly Driven Superconducting Qubits, 2005, Science 307 1210
[9] Gang Cao, Hai-Ou Li, Tao Tu, Li Wang, Cheng Zhou, Ming Xiao, Guang-Can Guo, Hong Wen Jiang&Guo-Ping Guo, Ultrafast universal quantum control of a quantum-dot charge qubits using Landau–Zener–Stückelberg interference, Nature Communication 4, 2013, 1401

[10] Mark G. Bason. Matthieu Viteau, Nicola Malossi *et. al.* High-fidelity quantum driving. Nature Physics 8, 2012, pp. 147-152

[11] Shuang Cong, Fangfang Meng , A Survey of Quantum Lyapunov Control Methods, The Scientific World Journal,Volume 2013, Paper ID 967529

[12] Koji Tsumura, Global Stablization at Arbitrary Eigenstates of N-dimensional Quantum Spin Systems via Continuous Feedback, 2008 American Control Conference, Washington, USA, June 11-13, 2008, pp. 4148-4153

[13] Shuang Cong, Longzhen Hu, Fei Yang, Jianxiu Liu, Characteristics Analysis and State Transfer for non-Markovian Open Quantum Systems, ACTA AUTOMATICA SINICA, 2013，30（4）, pp. 360-370

[14] Petta, J. R. et al. Coherent Manipulation of Coupled Electron Spins in Semiconductor Quantum Dots, Science 309, 2180 (2005)

[15] Wang Yao, Ren-Bao Liu, and L. J. Sham, Theory of electron spin decoherence by interacting nuclear spins in a quantum dot, Phys. Rev. B 74, 195301 (2006).

[16] G. Lindblad, On the Generators of Quantum Dynamical Semigroups Comm. Math. Phys. 48, 119-130 (1976).

[17] C. M. Quintana, K. D. Petersson, L. W. McFaul, S. J. Srinivasan, A. A. Houck, J. R. Petta, Cavity-mediated entanglement generation via Landau-Zener interferometry, Phys. Rev. Lett. **110**, 173603 (2013)

[18] L. S. Bishop, Circuit Quantum Electrodynamics Ph.D. thesis, Yale University (2010)